\documentstyle[emulateapj,apjfonts,psfig,times]{article}

\newcommand{\dpr}{$^{\prime\prime}\ $}

\slugcomment{Submitted to Astrophysical Journal}
\lefthead{Becker et al.}
\righthead{X-rays from the nearby solitary millisecond pulsar PSR J0030+0451}

\begin{document}

\title{X-RAYS FROM THE NEARBY SOLITARY MILLISECOND PULSAR PSR J0030+0451 --
THE FINAL ROSAT OBSERVATIONS}

\author{Werner Becker\altaffilmark{1}, 
        Joachim Tr\"umper\altaffilmark{1},
        Andrea N.~Lommen\altaffilmark{2},
        Donald C.~Backer\altaffilmark{2}} 

\altaffiltext{1}{Max-Planck-Institut f\"ur extraterrestrische Physik, 
 D-85740 Garching bei M\"unchen.} 
\altaffiltext{2}{Astronomy Department \& Radio Astronomy Laboratory 
 University of California, Berkeley, CA.} 

\begin{abstract}
  We report on X-ray observations of the solitary 4.8 ms pulsar PSR
  J0030+0451. The pulsar was one of the last targets observed in
  DEC-98 by the ROSAT PSPC. X-ray pulses are detected on a
  $4.5\sigma$ level and make the source the $11^{th}$ millisecond
  pulsar detected in the X-ray domain. The pulsed fraction is found
  to be $69\pm18\%$. The X-ray pulse profile is characterized by two
  narrow peaks which match the gross pulse profile observed
  at 1.4 GHz. Assuming a Crab-like spectrum the X-ray flux is in the
  range $f_x= 2-3\times 10^{-13}$ erg s$^{-1}$ cm$^{-2}\;$ ($0.1-2.4$
  keV), implying an X-ray efficiency of $L_x/\dot{E}\sim 0.5-5 \times
  10^{-3}\;(d/0.23 \mbox{kpc})^2$.
\end{abstract}


\section{INTRODUCTION} 

  In the $P-\dot{P}$ parameter space, millisecond pulsars are separate 
  from the majority of ordinary field pulsars. They are distinguished by 
  their short spin periods and small period derivatives, giving them high 
  spin-down ages of typically $10^9-10^{10}$ years and low magnetic field
  strengths of the order of $10^8 - 10^{10}$ G. More than $\sim 75\%$ of 
  the known disk millisecond pulsars are in binaries with a compact companion
  star, compared with the $\cong 1\%$ of binary pulsars found in the general 
  population. This gives support to the idea that their fast rotation has 
  been recycled by angular momentum transfer during a past mass accretion 
  phase (\cite{BisnovatyiKoganKomberg74} 1974; \cite{Alpar82}; 
  \cite{BhattacharyaHeuvel91} 1991; \cite{UrpinGeppertKonenkov98}1998). 
  Today, 95 recycled pulsars are known of which 57 are located in the 
  galactic plane (Camilo et al.~1999, Edwards et al.~2000, Lommen et al.~2000, 
  Lyne et al.~2000, Manchester et al.~2000). The others are in globular clusters 
  (Kulkarni \& Anderson 1996, Camilo et al.~2000) which provide a favorable
  environment for the recycling scenario (Rasio et al.~2000). Only 10 of the
  57 galactic millisecond pulsars are solitary (this includes PSR B1257+12
  which is in a planetary system);  the rest are in binaries 
  usually with a low-mass white dwarf companion (Camilo et al.~1999). The 
  formation of solitary recycled pulsars is not understood, 
  but it is widely believed that after being 'spun-up' in a binary system the 
  companion to the pulsar was either evaporated or the system was tidally 
  disrupted.

  Recycled pulsars were studied exclusively in the radio domain until 
  the early 1990's, when ROSAT, ASCA and BeppoSAX, which had significantly 
  higher sensitivities compared with previous observatories, were launched. 
  The first millisecond pulsar discovered as pulsating X-ray source by
  ROSAT was PSR J0437$-$4715 (Becker \& Tr\"umper, 1993). Further detections 
  which followed in recent years\footnote{For a review on the X-ray emission 
  properties and the observational status of recycled pulsars see Becker 
  \& Tr\"umper 1999 and Becker 2000.} sum up to almost $\sim 1/3$ 
  of all X-ray detected rotation-powered pulsars.

  The available data suggest that the observed emission is likely to be
  dominated by non-thermal processes. This is supported by observations
  of PSR B1821$-$24 (Kawai \& Saito 1999), PSR B1937+21 (Takahashi et al.~1999)
  and PSR J0218+4232 (Mineo et al.~2000) for which power-law spectra and
  pulse profiles with narrow peaks have been measured by ASCA, RXTE or 
  BSAX. For PSR J0437$-$4715 and PSR J2124$-$3358 the existing data do not 
  allow us to unambiguously discriminate between the two likely scenarios: 
  thermal emission from heated polar-caps or non-thermal emission from 
  within the co-rotating magnetosphere. All other X-ray detected recycled
  pulsars (PSRs B1957+20, J1012+5307, B0751+18, J1744$-$1134 and 
  J1024$-$0719) are identified only by their positional coincidence with 
  the radio pulsar and, in view of the low number of detected counts, do not 
  provide much more than a rough flux estimate. The power of XMM-Newton 
  and Chandra is needed to explore their emission properties in more detail. 
  However, the fact that all millisecond pulsars have roughly the same X-ray 
  efficiency ($L_X/\dot{E} \sim 10^{-3}$) as ordinary pulsars supports the 
  idea that the bulk of their X-ray emission has a common origin: magnetospheric, 
  viz.~non-thermal emission (Becker and Tr\"umper 1997).

  In this paper we report on soft X-ray observations of the solitary
  millisecond pulsar J0030+0451 using ROSAT. PSR J0030+0451 was discovered
  only recently with the Arecibo radio observatory during a drift scan search
  for pulsars (Lommen et al.~2000), and was detected independently during a search
  for sub-millisecond pulsars with the Bologna Northern Cross by D'Amico~(2000).
  The pulsar has a rotation period of $P=4.86$ ms and an apparent period 
  derivative of $\dot{P}=1.0(2) \times 10^{-20}\, \mbox{s s}^{-1}$ (Lommen et
  al.~2000). The intrinsic period derivative may be significantly smaller 
  than this due to the Shklovskii effect (Shklovskii et al.~1970; Camilo et 
  al.~1994). The pulsar spin-down age $P/2\dot{P} = 8 \times 10^9$ yrs is thus 
  a lower-limit and $B_\perp=2.2 \times 10^{8}$ G is an upper limit to the 
  surface magnetic field. The inferred rotational energy loss rate is $\dot{E}\le
  3.4\times 10^{33}\;\mbox{erg s}^{-1}\mbox{cm}^{-2}$. Using the model of 
  Taylor \& Cordes (1993) for the galactic distribution of free electrons, the 
  radio dispersion measure of $DM=4.33\; \mbox{pc cm}^{-3}$ implies a pulsar 
  distance of 230 pc and a column density of $N_H\sim 10^{20}\,\mbox{cm}^{-2}$.
  In terms of the pulsars spin-parameters PSR J0030+0451 thus turns out to 
  be a twin of the recent X-ray detected solitary millisecond pulsar 
  J2124$-$3358 (Becker \& Tr\"umper 1999; Bailes et al.~1997), making 
  it a very promising target for X-ray studies.

\section{OBSERVATION AND DATA ANALYSIS} 

 In the course of the ROSAT mission, PSR J0030+0451 was in the PSPC
 (Position Sensitive Proportional Counter) field of view on three
 different occasions. During the ROSAT All-Sky Survey (RASS) the pulsar
 was observed for 367 seconds between 1990 June 12-14, but the survey
 sensitivity of $\sim 3\times 10^{-12}\,\mbox{erg s}^{-1}\mbox{cm}^{-2}$
 was not sufficient to detect the pulsar. A search for serendipitous
 data in the ROSAT archive showed that the pulsar position was observed
 in 1992 July at about $\sim 43$ arcmin off-axis. Unfortunately, the
 pulsar position turned out to be fully covered by a laceration in the PSPC
 entrance window support structure, reducing the detection sensitivity
 for the pulsar in that data to almost zero. When it became clear
 that the ROSAT mission would come to its ultimate end the PSPC detector
 was activated again after being out of the focus for more than two years.
 Among a few other sources, PSR J0030+0451 was scheduled to exhaust the
 remaining detector gas before the satellite was switched off. The
 observations took place on 1998 December 16-17 and provided us
 with 7743 sec of good data. The pulsar, together with the quasar
 GB1428+4217 (Boller et al.~2000), were the final targets observed by
 ROSAT.

 The unstable behavior of the satellite during its last observations
 as well as the reduced gas flow and pressure in the PSPC caused
 several data anomalies. The reduced and non-uniform gas flow led 
 to a variable and somewhat lower gain. Although the effect 
 of this gain variation has been corrected during the standard 
 processing, some soft events may have escaped detection due to 
 their falling below the raw amplitude lower limit, i.e.~event 
 numbers in the channels below 15 (corresponding to about 0.15 keV) are 
 incomplete.  We have corrected for occasional master veto rates 
 below $35\,\mbox{counts s}^{-1}$ which indicate occasional breakdowns 
 of the high voltage not recorded in the housekeeping data. In addition, 
 we corrected for an occasional flickering of the high voltage which 
 caused periods of frequent short accepted time intervals of less than 
 10s where the nominal voltage (and thus gain value) may not have been 
 reached.
 Finally, a ``sensitivity hole'' appeared to develop in the 
 north-western quadrant of the PSPC detector, which reached 
 the on-axis position at the  time of our observation.  It is
 not possible to correct for this degradation in signal.
 Luckily, the millisecond pulsar position available at the time
 of scheduling was offset in right ascension and declination by
 a few arcmin in a direction opposite from the hole. 

\subsection{Spatial analysis}
\begin{figure}
 \centerline{\psfig{figure=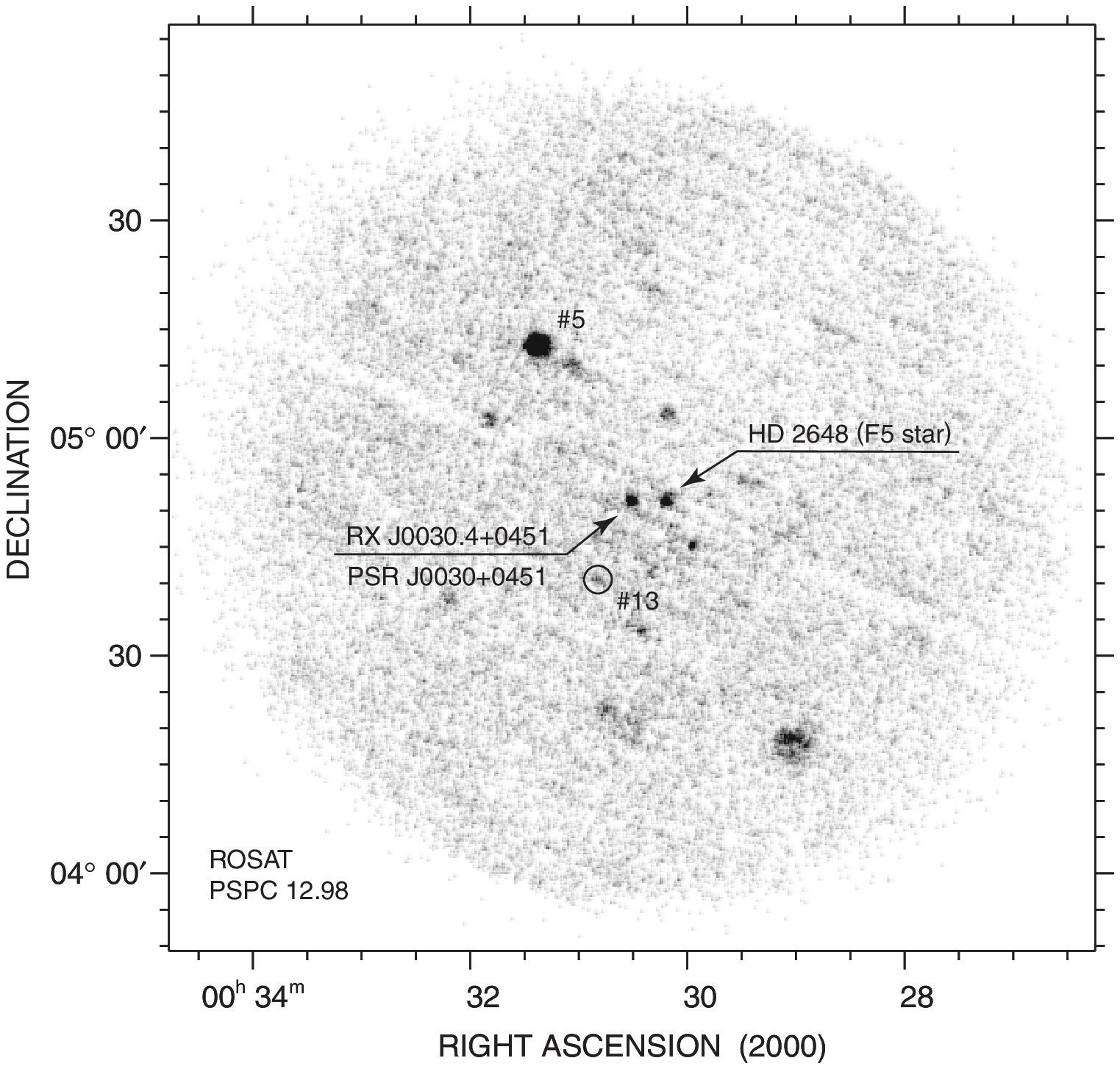,width=15cm}}
 \caption[]{The two degree ROSAT PSPC image of the field around
  PSR J0030+0451. RX J0030.4+0451 and the F5-star HD 2648 are
  located close to the optical axis, with a separation of $\sim
  4$ arcmin. The sensitivity hole is only barely visible in the
  north-western quadrant of the image.  The two sources which are
  coincident with NVSS radio sources are indicated.}
  \label{PSPC_image}
\end{figure}

 We searched for the X-ray counterpart of PSR J0030+0451 by doing
 a maximum likelihood analysis of the source counts in combination 
 with a spline fit to the background level and fitting the results.
 After applying the data corrections described above, 17 X-ray 
 sources were detected in the PSPC field of view with a detection 
 threshold of $\ge 5\sigma$ (cf. Fig.\ref{PSPC_image}). The sources
 are listed in Table \ref{source_list}. Because of the sensitivity 
 in the north-western part of the detector the source list might 
 not be complete for that sky  region.

 \placetable{source_list}

 The loss of the last ROSAT star tracker in April 1998, which
 was substituted by the Wide-field camera star tracker, caused
 systematic position inaccuracies which were as large as 30 arcsec.
 Correlating the ROSAT sources with optical catalogues we find
 that RX J0030.1+0451 is within $\sim 17$ arcsec with HD 2648,
 an F5-star with m$_v=6.7$. This source is also detected in the
 ROSAT all-sky survey data and in the serendipitous archival data, 
 although in the latter at an off-axis angle of $\sim 40$ arcmin 
 and close to the PSPC entrance window support structure. Its 
 position accuracy given in the WGA catalog (White et al.~1994)
 is, for that reason, not very accurate. The position determined 
 in the RASS data is about 1.5 arcmin off
 from the position obtained in the December 98 data. The X-ray 
 emission is very soft with practically no emission beyond 0.5 keV.

 \placefigure{PSPC_image}

 Accepting the identification of RX J0030.1+0451 with HD 2648 we
 are able to use its optical position to improve the positional
 accuracy of all X-ray sources in the field. The differences
 in right ascension and declination for HD 2648 are found to be
 RA$_{opt}-$RA$_{Rosat}$=0.45\dpr and DEC$_{opt}-$DEC$_{Rosat}
 =-16.2$\dpr. Applying these corrections to all X-ray sources 
 resulted in the positions given in Table \ref{source_list}.
 Correlating the millisecond pulsar position  with the corrected 
 source locations, we find that RX J0030.4+0451 is within $\sim 11$ 
 arcsec to the radio position of PSR J0030+0451, making it a likely
 counterpart for the millisecond pulsar. Additionally, the USNO catalog  
 lists an optical source with $m_V=18.3$ and $m_B=19.5$ at 
 RA=00:30:27.84 and DEC=+04:51:34.74, which is only about 6 arcsec 
 distant from the millisecond pulsar position. Although the USNO 
 source may be the optical counterpart of the millisecond pulsar 
 (which is expected to be much fainter at optical wavelength than 
 $m_V\approx 19$), it is not a priori excluded that it is the 
 counterpart of the ROSAT source RX J0030.4+0451 rather than the 
 millisecond pulsar. With\,$B-V=1.2$ the optical object could be 
 a K5 star at $\sim 1.5$ kpc, but an AGN is also possible. For 
 $\log f_x/f_v$ we determine a value close to zero, which 
 according to Maccacaro et al.~(1988) would be in agreement 
 with an AGN. The identification of a timing signature according 
 to the millisecond pulsar rotation period is therefore necessary
 to unambiguously identify RX J0030.4+0451 as being the X-ray 
 counterpart of PSR J0030+0451.

 Two other X-ray detected sources in the PSPC field were identified
 with sources in the NRAO VLA Sky Survey at 1.4 GHz (NVSS, Condon et
 al.~1998) to within 10$^{\prime\prime}$ and are shown in Table
 \ref{source_list}.  This survey has a completeness limit of about
 2.5 mJy.  There are no NVSS sources within 3$^\prime$ of any other
 PSPC source in this field.

\subsection{Timing analysis}

 In order to search for a 4.86 ms spin modulation from RX J0030.4+0451
 we have selected all photons within an annulus of 70 arcsec centered
 on the source. The selection radius includes 103 events of which
 $\sim 10\%$ belong to the background. The X-ray arrival times were
 corrected for the satellite's motion, and a barycenter correction
 was performed using the analysis software {\em eXsas} (Zimmermann
 et al.~1997). For the arrival time correction we fitted the clock
 calibration points, available for the days $334-340$ of 1998, with
 a first order polynomial\footnote{Our fit is included in the ROSAT
 clock calibration table which is distributed as part of {\em eXsas}
 version APR-2000.}.

 Millisecond pulsars are stable clocks. Given the high precision of the
 pulsar's radio ephemeris (cf.~Table \ref{ephemeris}), we related each
 X-ray photon arrival time directly to the pulsar's rotation phase
 $\phi$. By applying the $Z^2_n-$test with $n=1$ to $10$ harmonics
 (Buccheri \& De Jager 1989) in combination with the H-Test, to
 yield the optimal number of harmonics for a pulsed signal
 (De Jager 1987), we find a deviation from a flat pulse-phase
 distribution with a significance of 4.5$\sigma$ for 4 harmonics.
 This establishes RX J0030.4+0451 as the X-ray counterpart
 of PSR J0030+0451 beyond any doubt, and makes it the 11$^{th}$
 millisecond pulsar detected in the soft X-ray domain.

 \placetable{ephemeris}

\begin{figure}
 \centerline{\psfig{figure=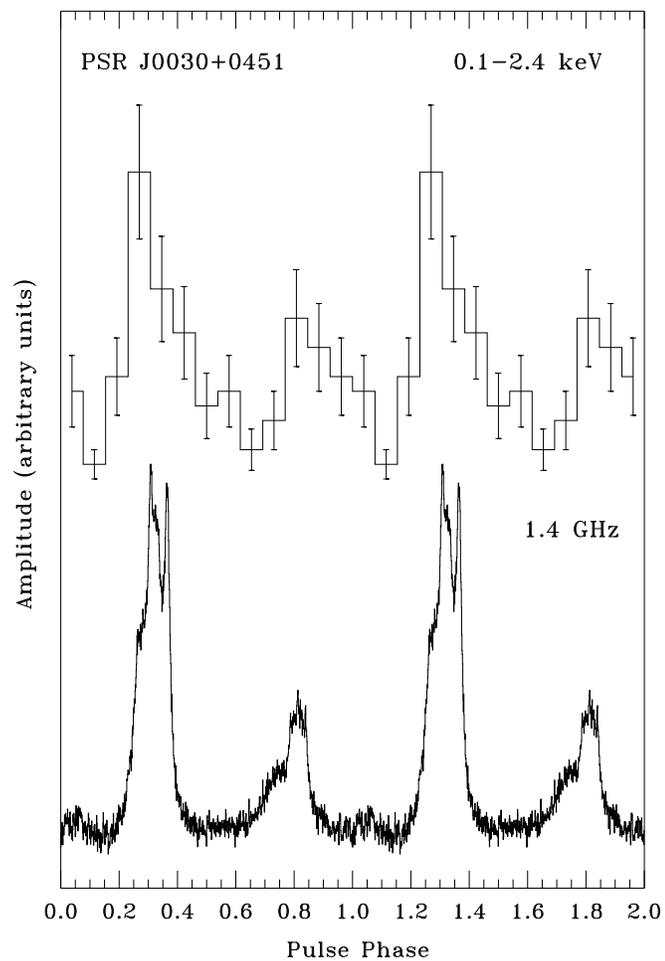,width=10cm}}
 \caption[]{X-ray and radio pulse profile of PSR J0030+0451 as observed
  with the ROSAT PSPC in the 0.1$-$2.4 keV band (top) and the Arecibo radio
  telescope at 1.4 MHz (bottom). Two phase cycles are shown for clarity.
  The phase alignment is arbitrary.}
 \label{pulse_profile}
\end{figure}

 The X-ray pulse profile of PSR J0030+0451 is shown in
 Figure \ref{pulse_profile}. It is characterized by two narrow peaks
 that are separated in phase by $\sim 180^\circ$. The fraction of pulsed 
 photons, which was determined by the bootstrap method described in 
 Becker \& Tr\"umper (1999), is $69\pm 18 \%$. Fig.\ref{pulse_profile} 
 includes the radio pulse profile at 1.4 GHz for comparison with arbitrary
 phase alignment owing to the lack of absolute timing on ROSAT. The gross 
 pulse morphology is the same in the two observing bands. The fine structure
 seen in the radio profile is not resolved in the X-ray profile owing 
 to the limited photon statistics. 

 \placefigure{pulse_profile}

\subsection{Spectral analysis}
  The low number of detected source counts together with the unstable
  behavior of the PSPC detector during the DEC-98 observations does not
  support a detailed spectral analysis of the pulsar data. We
  estimated the pulsar's energy output in the ROSAT band by
  converting the PSPC source count rate into an energy flux
  assuming a power-law spectrum $dN/dE \propto E^{-\alpha}$ with
  photon-index $\alpha=2$ (c.f.~Becker \& Tr\"umper 1997). The column
  density along the pulsar's line of sight is $N_H(galaxy)= 3\times 10^{20}\,
  \mbox{cm}^{-2}$ through the Galaxy (Dickey \&  Lockman 1990).
  Assuming an  electron density of $n_e=0.03\; \mbox{cm}^{-3}$  the
  radio dispersion measure converts to $N_H(pulsar)= 1.3\times 10^{20}\,
  \mbox{cm}^{-2}$. Both numbers are of the same order and yield
  fluxes in the range  $f_x=(2-3) \times 10^{-13}$ erg s$^{-1}$
  cm$^{-2}\;$ ($0.1-2.4$ keV). The corresponding (isotropic) X-ray
  luminosity is $L_x \sim (1-2) \times 10^{30}\;(d/0.23 \mbox{kpc})^2$
  erg s$^{-1}$. As mentioned already, the Shklovskii contribution
  to the pulsar's period derivative is unknown.  Assuming a typical
  pulsar space velocity of 65 km s$^{-1}$, the intrinsic $\dot{P}$
  of PSR J0030+0451 would be an order of magnitude smaller than
  what is measured.  Each parameter derived from $\dot{P}$ also
  inherits the same possibility for correction. In particular, the
  rotational energy loss rate of $\dot{E}=3.4\times 10^{33}\,
  \mbox{erg s}^{-1}$ is an upper limit and could be a factor
  of 10 smaller when the proper motion is measured. Taking this
  uncertainty into account the pulsar's X-ray efficiency is
  in the range $\eta = L_x/\dot{E} \sim (0.5-5) \times 10^{-3}\;
  (d/0.23 \mbox{kpc})^2$.

\section{Summary and Conclusion}

  PSR J0030+0451 is the $11^{th}$ millisecond pulsar detected in
  the soft X-ray band. In terms of its spin-parameters, it is
  almost a ``twin'' of the solitary millisecond pulsar J2124$-$3358
  from which pulsed X-ray emission was discovered in a recent
  HRI observation (Becker \& Tr\"umper 1999). Although the
  present data do not support a detailed spectral analysis, the
  temporal emission characteristic of a high pulsed fraction
  ($\sim 69\pm 18\%$), the phase separation of $\sim 180^\circ$ between
  the two narrow  peaks, and the gross similarity  between the radio
  and X-ray pulse profile suggest that the mechanisms which create
  the pulsed X-rays are closely related to the radio emission process.
  The bulk of the observed X-rays is then likely to be of non-thermal
  origin.

  It has been recently argued that the group of X-ray detected ms-pulsars
  can be divided into two classes (see e.g.~Kawai \& Saito 1999). The first 
  includes PSR B1821$-$24, B1937+21 and J0218+4232 ($P\sim 1.5-3$ ms, 
  $\log\dot{E}\sim 35-36$ erg s$^{-1}$), for which the X-ray emission 
  exhibits power-law spectra and the pulse profiles have narrow peaks, 
  and therefore magnetospheric emission is invoked. The non-thermal, 
  magnetospheric emission is expected to have a power-law energy spectrum.
  The second class comprises PSR J0437$-$4715 and J2124$-$3358 ($P\ge 5$ ms,
  $\log\dot{E}\sim 33-34$ erg s$^{-1}$) for which the X-ray emission is
  believed to be dominated by thermal polar-cap emission. Unfortunately
  the  existing spectral data are not of sufficient quality to discriminate
  between the thermal polar-cap and non-thermal magnetospheric emission
  scenarios. Both models fit the data equally well. The argument for
  this second group being thermal emitters is that their pulse profiles
  are broad, and the pulsed fraction is small (pulsed fraction $\sim
  20-30\%$) when compared to the non-thermal emitting sources (pulsed
  fraction $\ge 50-60\%$). The argument is not unreasonable --  thermal
  polar-cap emission should result in broad sinusoidal soft modulated
  emission with a pulsed fraction $\le 50\%$ (Pavlov \& Zavlin 1998),
  whereas magnetospheric emission is reasonably expected to cause
  narrow pulse-peaks and large pulsed fractions.

  Despite the apparent reasonableness of the previous arguments, the
  interpretation is not unique. Becker \& Tr\"umper (1997; 1999)
  have emphasized that the radiation cone produced by magnetospheric
  emission, which yields sharp peaks at one aspect angle, may well be
  less sharply modulated when viewed from other directions. Thus a broad,
  weakly modulated pulse profile is not unambiguous evidence for thermal
  emission. Indeed, one can argue on phenomenological grounds based
  on the $L_x$ vs $\dot{E}$ relation (most of the X-ray pulsars lie close
  to the line $L_x \sim 10^{-3} \dot{E}$) that the X-ray emission from
  all ms-pulsars is dominated by magnetospheric emission. One could
  invoke a gross similarity between the X-ray and radio pulse profiles
  as part of this argument  (see Fig. 8 and Fig. 13 from Becker \& 
  Tr\"umper 1999). It would seem unlikely that the physical
  conditions involved change so dramatically with the parameters as to
  produce two distinct classes of sources. There are reasonable physical
  models that allow the radio and the X-ray  emission to be produced by
  processes taking place in the pulsar magnetosphere from which the
  radio emission arises (cf.~Crusius-W\"atzel, Kunzel \& Lesch 2000).
  There can still be thermal emission from the polar caps, e.g.,~heating
  by particle back-flow from the magnetosphere and non-thermal
  magnetospheric emission from charged particles moving out along
  the curved open field lines, but the key question is how both
  emission mechanisms add together (e.g.~Zhang \& Harding 2000).

  PSR J0030+0451 thus is found to be a hybrid. While its spin 
  parameters suggest smoothly pulsed thermal emission, its 
  observed X-ray pulse profile and pulsed fraction suggest a 
  non-thermal origin of the X-ray emission. So have we fooled 
  ourselves into creating two classes of millisecond pulsars? 
  Is it possible that the emission from all recycled pulsars 
  is simply due to a gradual variation of the parameters defining 
  the magnetosphere, or will we need another interpretation entirely? 
  Further X-ray observations by Chandra or XMM will surely shed 
  some light on this topic providing detailed spectral and temporal
  information from this peculiar object.

\begin{acknowledgements}
  The ROSAT project is supported by the Bundesministerium f\"ur Bildung,
  Wissenschaft, Forschung und Technologie (BMBF) and the Max-Planck-Society
  (MPG). We thank our colleagues from the MPE ROSAT group for their support
  and M.~Freyberg for providing us with the EXSAS routines required to
  correct the PSPC anomalies present in the DEC 98 data. We also wish to
  acknowledge discussions with M.C. Weisskopf and thank the anonymous referee
  for his comments.
\end{acknowledgements}

\newpage

 \begin{deluxetable}{c c c c c c c c c}
 \tablewidth{40pc}
 \tablecaption{X-ray sources detected with a significance of
 $\ge 5\sigma$ within the ROSAT PSPC's field of view during
 the Dec 1998 observation of PSR J0030+04551. \label{source_list}}
 \tablehead{
 \multicolumn{6}{c}{PSPC source detected}  & \multicolumn{3}{c}{NVSS source or Pulsar detected} \\
 \cline{2-5}\cline{7-9} \\
 \colhead{Nr.} & \colhead{Name} & \colhead{RA(2000)} & \colhead{DEC(2000)} &
 \colhead{Rate} & \colhead{Error} & \colhead{RA(2000)} & \colhead{DEC(2000)} & \colhead{Flux}\\ 
 \colhead{} & \colhead{RX} & \colhead{HMS} &
 \colhead{DMS} & \colhead{cts/s} & \colhead{cts/s} & \colhead{HMS} & \colhead{DMS} & \colhead{mJy}}
 \startdata
   1 & J0030.3+0527  & 00,30,23.86 & +05,27,36.2 & 0.0095 & 0.002 & & & \\
   2 & J0030.2+0521  & 00,30,16.90 & +05,21,32.6 & 0.0077 & 0.002 & & & \\
   3 & J0032.9+0518  & 00,32,59.39 & +05,18,22.7 & 0.0175 & 0.003 & & & \\
   4 & J0028.6+0516  & 00,28,36.60 & +05,15,48.4 & 0.0127 & 0.002 & & & \\
   5 & J0031.3+0513  & 00,31,20.93 & +05,13,26.7 & 0.1189 & 0.005 & 00,31,20.87 & +05,13,21.6 & 8.8 \\
   6 & J0031.0+0511  & 00,31,01.65 & +05,10,45.1 & 0.0125 & 0.002 & & & \\
   7 & J0030.1+0504  & 00,30,08.11 & +05,03,53.9 & 0.0084 & 0.001 & & & \\
   8 & J0031.7+0503  & 00,31,47.85 & +05,03,03.5 & 0.0097 & 0.002 & & & \\
   9 & J0029.4+0454  & 00,29,25.63 & +04,54,31.1 & 0.0034 & 0.001 & & & \\
 PSR & J0030.4+0451  & 00,30,28.14 & +04,51,41.4 & 0.0156 & 0.002 & 00,30,27.43 & +04,51,39.7 & 0.6 \\
  11 & J0030.1+0451  & 00,30,08.74 & +04,51,36.7 & 0.0338 & 0.009 & & & \\
  12 & J0029.8+0445  & 00,29,53.87 & +04,45,22.4 & 0.0076 & 0.001 & & & \\
  13 & J0030.7+0440  & 00,30,47.38 & +04,40,41.7 & 0.0039 & 0.001 & 00,30,47.71 & +04,40,38.3 & 4.1 \nl
  14 & J0032.1+0438  & 00,32,10.80 & +04,37,50.5 & 0.0090 & 0.002 & & & \\
  15 & J0030.3+0433  & 00,30,22.83 & +04,33,26.7 & 0.0058 & 0.001 & & & \\
  16 & J0030.5+0421  & 00,30,35.26 & +04,21,04.9 & 0.0297 & 0.003 & & & \\
  17 & J0028.9+0417  & 00,28,57.01 & +04,17,37.3 & 0.0839 & 0.005 & & & \\
 \enddata
 \tablecomments{The count rate given for RX J0030.1+0451 (HD 2648)
 was taken from the ROSAT all-sky survey data as this source turned
 out to be strongly effected by the PSPCs sensitivity hole.}
 \end{deluxetable}

\clearpage

  \begin{deluxetable}{l l}
   \tablewidth{0pc}
   \tablecaption{Pulsar Parameters for PSR J0030+0451 \label{ephemeris}} 
    \tablehead{}
   \startdata 
     Right ascension (J2000)      & $00^h\; 30^m\; 27^s\!.432$       \\
     Declination (J2000)          & $+04^\circ\; 51'\; 39"\!.7$      \\
     Pulsar Period, P             & 0.00486545320737  ms             \\
     Period derivative            & $1.0 \times 10^{-20}$            \\
     Epoch of period, $T_0$       & MJD 50984.4                      \\
     Dispersion Measure           & DM 4.33 pc cm$^{-3}$             \\
     Dispersion based distance    & d = 230 pc                       \\
   \enddata
   \tablecomments{From Lommen et al.~2000}
   \end{deluxetable}

\clearpage

\clearpage

\begin{thebibliography}{}

\bibitem[Alpar et al.~1982]{Alpar82}
 Alpar M.A., Cheng A.F., Ruderman M.A., Shaham J., 1982, Nat, 300, 728

\bibitem[Bailes et al.~]{Bailes97}
Bailes, M., Johnston, S., Bell, J.~F., Lorimer, D.~R., Stappers, B.~W.,
  Manchester, R.~N., Lyne, A.~G., Nicastro, L., D'Amico, N., \& Gaensler, B.~M.
  1997, ApJ, 481, 386

\bibitem[Becker \& Tr\"umper~]{BeckerTrumper93}
 Becker, W., \& Tr\"umper, J. 1993, Nat, 365, 528

\bibitem[Becker \& Tr\"umper~]{BeckerTrumper97}
 Becker, W., \& Tr\"umper, J. 1997, A\&A 326, 682

\bibitem[Becker \& Tr\"umper~]{BeckerTrumper99}
 Becker, W., \& Tr\"umper, J. 1999, A\&A, 341, 803

\bibitem[Becker ]{Becker2000}
 Becker, W. 2000, Astro. Lett. \& Comm., in press

\bibitem[Bhattacharya \& Van den Heuvel]{BhattacharyaHeuvel91}
 Bhattacharya D., Van den Heuvel E.P.J., 1991, Phys.~Rep., 203, 1

\bibitem[Bisnovatyi-Kogan \& Komberg]{BisnovatyiKoganKomberg74}
 Bisnovatyi-Kogan G.S., Komberg B.V., 1974, SvA, 18, 217

\bibitem[Boller et al.~]{Boller_et_all_2000}
 Boller, T., Fabian, A.C., Brandt, W.N., \& Freyberg, M.J., 2000, MNRAS, in press

\bibitem[Buccheri \& De Jager]{BuccheriDeJager89}
 Buccheri, R., \& De Jager, O.C. 1989, in {\it Timing Neutron Stars}, ed. H.\"Ogelman,
 E.P.J.~van den Heuvel, [Dordrecht : Kluwer], p. 95

\bibitem[Camilo et al.]{Camilo94}
 Camilo, F., Thorsett, S.E., Kulkarni, S.R., 1994, ApJ, 421, L15

\bibitem[{Camilo}(1999){Camilo}]{Camilo99}
{Camilo}, F. 1999, In {\em Pulsar Timing, General Relativity and the Internal
  Structure of Neutron Stars}, ed. Z. Arzoumanian, F.~Van der Hooft, \&
  E.P.J.~van den Heuvel, [Amsterdam:Koninklijke Nederlandse Akademie van
  Wetenschappen], p. 115

\bibitem[{Camilo} {et~al.}(2000){Camilo}, {Lorimer}, {Freire}, {Lyne}, and
  {Manchester}]{clf+00}
{Camilo}, F., {Lorimer}, D.~R., {Freire}, P., {Lyne}, A.~G., \& {Manchester},
  R.~N. 2000, \apj, in press (astro-ph/9911234)

\bibitem[Condon {et~al.}(1998)Condon, Cotton, Greisen, and Yin]{Condon98}
Condon, J.~J., Cotton, W.~D., Greisen, E.~W., \& Yin, Q.~F. 1998, AJ, 115, 1693

\bibitem[Crusius-W\"atzel et al.]{Crusius}
 Crusius-W\"atzel, A., Kunzel, T., Lesch, H., 2000, ApJ, submitted

\bibitem[Kulkarni \& Anderson(1996)Kulkarni and Anderson]{Kulkarni96}
Kulkarni, S.~R., \& Anderson, S.~B. 1996, in {\em {D}ynamical {E}volution of {S}tar
  {C}lusters -- {C}onfrontation of {T}heory and {O}bservations}: {IAU}
  {S}ymposium 174, Kluwer Academic Publisher, p.181
  
\bibitem[Lyne {et~al.}(2000)Lyne, Camilo, Manchester, Bell, Kaspi, D'Amico,
  McKay, Crawford, Morris, Sheppard, and Stairs]{Lyne00}
  Lyne, A.~G., Camilo, F., Manchester, R.~N., Bell, J.~F., Kaspi, V.~M., D'Amico,
  N., McKay, N. P.~F., Crawford, F., Morris, D.~J., Sheppard, D.~C., \& Stairs,
  I.~H. 2000, \mnras, in press (astro-ph/9911313)

\bibitem[{Manchester} {et~al.}(2000){Manchester}, {Lyne}, {Camilo}, {Kaspi},
  {Stairs}, {Crawford}, {Morris}, {Bell}, and {D'Amico}]{Manchester00}
  {Manchester}, R.~N., {Lyne}, A.~G., {Camilo}, F., {Kaspi}, V.~M., {Stairs},
  I.~H., {Crawford}, F., {Morris}, D.~J., {Bell}, J.~F., \& {D'Amico}, N.,
   2000, in {\it Pulsar Astronomy - 2000 and Beyond}, 
   ed.  M. Kramer, N. Wex, and R. Wielebinski, [San Francisco : ASP], p. 49

\bibitem[{Edwards}(2000){Edwards}]{Edwards00}
  {Edwards}, R.~T. 
   2000, in {\it Pulsar Astronomy - 2000 and Beyond}, 
   ed.  M. Kramer, N. Wex, and R. Wielebinski, [San Francisco : ASP], p. 33 

\bibitem[D'Amico]{Amico00}
 D'Amico, N. 
   2000, in {\it Pulsar Astronomy - 2000 and Beyond}, 
   ed.  M. Kramer, N. Wex, \& R. Wielebinski, [San Francisco : ASP], p. 27 

\bibitem[De Jager]{DeJager87}
 De Jager, O.C. 1987, PhD Thesis, Potchefstromm University for
 Christian Higher Education, South Africa

\bibitem[Dickey \& Lockman]{DickeyLockman90}
 Dickey, J.M., Lockman, \& F.J., 1990, ARA\&A, 28, 215

\bibitem[Kawai \& Saito]{KawaiSaito99}
 Kawai, N., \& Saito, Y. 1999, Astro.~Lett. and Communications, 38, 1

\bibitem[Lommen et al.~]{Lommen00}
 Lommen, A.N., Zepka, A., Backer, D.C., Cordes, J.M., Arzoumanian, Z.,
 McLaughlin, M., \& Xilouris, K. 2000, ApJ, submitted

\bibitem[Maccacaro et al.~]{Maccacaro}
 Maccacaro, T., Isabella, M.G., Wolter, A., Zamorani, G., \& Stocke, J.T.
 1988, ApJ, 326, 680

\bibitem[Mineo et al.~]{Mineo2000}
 Mineo, T., Cusumano, G., Kuiper, L., Hermsen, W., Massaro, E., Becker, W., 
 Nicastro, L., Sacco, B., Verbunt, F., Lyne, A.G., Stairs, I.H., \& Shibata, S. 
 2000, A\&A, in press

\bibitem[Rasio et al.~]{Rasio}
 Rasio, F.A., Pfahl, E.D., \& Rappaport, S. 2000, ApJ, 532, L47

\bibitem[Shklovskii 1970]{Shklovskii70}
 Shklovskii, I., 1970, Soviet Astron., 13, 562

\bibitem[Takahashi et al.~]{Takahashi99}
 Takahashi, M.,  Shibata, S.,  Gunji, S.,  Sakurai, H., Torri, K., Saito, Y., 
 Kawai, N., Dotani, T., \& Hirayama, M. 1999, Astron.Nachr., 320, 340

\bibitem[Taylor \& Cordes]{TaylorCordes93}
 Taylor, J.H., \& Cordes, J.M. 1993, ApJ, 411, 674

\bibitem[Urpin et al.~]{UrpinGeppertKonenkov98}
 Urpin V., Geppert U., Konenkov D., 1998, A \& A, 331, 244

\bibitem[White et al.~]{WGA94}
White, N.E., Giommi, P., Angelini, L., 1994, IAU Circular 6100

\bibitem[Zavlin \& Pavlov]{Zavlin98}
 Zavlin, V.E., \& Pavlov, G.G. 1998, A\&A, 329, 583

\bibitem[Zhang \& Harding]{Zhang00}
 Zhang, B., \& Harding, A.K. 2000, ApJ, in press (astro-ph/9911028)

\bibitem[Zimmermann et al.~]{ZimmermannBeckerBelloni94}
 Zimmermann, H.U., Becker, W., Belloni, T., D\"obereiner, S., 
 Izzo, C., Kahabka, P., Schwentker O., et al. 1994, MPE-Report 244

\end{thebibliography}
\end{document}